\documentclass[aps,nofootinbib,twocolumn,floatfix]{revtex4}
\usepackage{graphicx}
\graphicspath{{figures/}} 
\usepackage{amsmath}
\usepackage{amsfonts}
\usepackage{amssymb}
\usepackage{dcolumn}
\usepackage{braket}
\usepackage{commath}
\usepackage{epstopdf}
\usepackage{tabularx, booktabs}
\usepackage[labelsep=quad,indention=0pt]{subfig}
\usepackage{color} 

\def\comment#1{}

\begin{document}

\title{The Kronig-Penney model extended to arbitrary potentials via numerical matrix mechanics}
\author{R.~L.~Pavelich}
\email{rpavelic@ualberta.ca}
\author{F.~Marsiglio}
\email{fm3@ualberta.ca}
\affiliation{Department of Physics, University of Alberta, Edmonton, AB, Canada T6G 2G7}

\begin{abstract}
\noindent{The Kronig-Penney model describes what happens to electron states when a confining potential is
repeated indefinitely. This model uses a square well potential; the energies and eigenstates can be obtained
analytically for a the single well, and then Bloch's Theorem allows one to extend these solutions to the periodically repeating square well potential. In this work we describe how to obtain simple numerical solutions for the eigenvalues and eigenstates for any confining potential within a unit cell, and then extend this procedure, with virtually no extra effort, to the case of periodically repeating potentials. In this way one can study the band structure effects which arise from differently-shaped potentials. One of these effects is the electron-hole mass asymmetry. More realistic unit cell potentials generally give rise to higher electron-hole mass asymmetries.}
\end{abstract}

\date{\today} 
\maketitle


\section{INTRODUCTION}

\noindent The Kronig-Penney model \cite{kronig31} remains the paradigm for the demonstration of energy bands, separated
by gaps, in periodic solids. This in turn is used to introduce the notion of metals vs.~(band) insulators, dependent on the position
of the chemical potential (and therefore dependent on the number of electrons) in the solid. The model is appealing insofar
as it only requires solutions to the Schr\"odinger equation for the simplest potential (a constant), and the equation for determining
allowed energies can be written analytically in terms of elementary functions. This exercise is also enlightening as it puts
into practice what might be at first viewed as a somewhat abstract theorem (Bloch's Theorem), and it also illustrates how
boundary conditions are paramount in determining the energy spectrum. The energies themselves are readily determined
through an explicit equation, and can be done with a calculator. Even more importantly, perhaps, is that in one particular limit
of the model, the so-called Dirac comb, where the barriers are made to be infinitely thin while their heights are made infinitely
high (i.e. a series of periodic $\delta$-functions), the existence of energy gaps can be demonstrated analytically.

All of these features are beneficial and even desirable for pedagogical reasons, but often one would like to demonstrate the
same ideas for more complicated (and realistic) potentials. In other words, it naturally occurs to students at some point to inquire about the generality of the conclusions reached through a model that considers only the simplest possible case. Furthermore, 
for graduate students there tends to be a quantum leap from the analytical solution of the Kronig-Penny model to the numerical
(black box-like) solutions inherent in electronic band structure calculations. In this paper we propose to bridge this gap somewhat with simple, albeit numerical, solutions for an arbitrary unit cell potential, though still in one dimension.

The basic idea comes from some recent work concerning very simple one-body potentials.\cite{marsiglio09,jelic12} We will
describe this work in the next section, and follow it with the alterations necessary to utilize periodic (as opposed to open)
boundary conditions. The purpose of this change is it allows us to follow up with an implementation of the Bloch condition, 
as opposed to the periodic boundary condition. We can then compare our numerically derived results with analytic ones,
to establish the validity and accuracy of the method. Note that an alternate procedure would be to simply define a
periodic potential that repeats a finite number of times, say ten or twelve, and solve the Schr\"odinger equation for this
potential. This was done, for example, in Ref. [\onlinecite{johnston92}], by solving the differential equation using a `hunt
and shoot' method. Here we prefer to use the matrix diagonalization method,\cite{marsiglio09} and we prefer to incorporate
Bloch's Theorem analytically, for the sake of efficiency and pedagogy. In the following section we explore a number 
of fairly different looking unit cell potentials; freedom to choose the parameters that characterize different shaped potentials makes each potential able to have various bandwidths and effective masses. One particular feature that is shape-dependent is the asymmetry in electron vs.~hole effective mass. We conclude with a summary.

\section{FORMALISM}

\subsection{Infinite square well}

\noindent{By embedding a potential in the infinite square well, one is able to study the bound states of that potential with minimal
technical overhead. One has to have access to a numerical matrix diagonalization routine; otherwise only a freshman
knowledge of integral calculus and linear algebra is required. The methodology is outlined in Ref.~\onlinecite{marsiglio09}
so only key points will be recalled here. One first expands the wave function in the infinite well basis:}
\begin{equation}
\ket{\psi} = \sum_{m=1}^{\infty} c_m \ket{\psi_m}
\label{expansion}
\end{equation}
\noindent where the basis states are
\begin{equation}
\psi_n(x) = \begin{cases}
\sqrt{\frac{2}{a}} \sin{\left( \frac{n \pi }{a}x \right)} & 0<x<a, \\
0 & \text{otherwise,}
\end{cases}
\label{eq:isweigvec}
\end{equation}
with eigenvalues
\begin{equation}
E_n^{(0)} = \frac{n^2 \pi^2 \hbar^2}{2ma^2} \equiv n^2 E_1^{(0)},
\end{equation}
\noindent where $a$ is the width of the well, and the (unknown) coefficients are $c_n$.
Straightforward algebra leads to the matrix diagonalization problem,
\begin{equation}
\sum_{m=1}^{\infty}  H_{nm} c_m = E c_n
\label{mat}
\end{equation}
where
\begin{eqnarray}\nonumber
H_{nm} &=& \bra{\psi_n} (H_0 + V) \ket{\psi_m} \\
&=& \delta_{nm} E_{n}^{(0)} + H_{nm}^{V}
\label{eq:hnmv}
\end{eqnarray}
and 
\begin{eqnarray}
H_{nm}^{V} &= &\bra{\psi_n} V(x) \ket{\psi_m} \nonumber \\
                     & = & {2 \over a} \int_0^a \ \dif x \ \sin{\left({n \pi x \over a}\right)} V(x) \sin{\left({m \pi x \over a}\right)}.
\label{matrix_ele}
\end{eqnarray}
The problem is easily rendered into a convenient dimensionless form by using the infinite square well width $a$ and
ground state energy $E_1^{(0)}$:
\begin{equation}
\sum_{m=1}^{\infty}  h_{nm} c_m = e c_n,
\label{mat_dim}
\end{equation}
where $h_{nm} \equiv H_{nm}/E_1^{(0)}$ and $e \equiv E/E_1^{(0)}$.

\subsection{Square well with periodic boundary conditions}

\noindent{In this work, as a first step, we use the same strategy, but instead of a box with infinite walls, we will use a box with
periodic boundary conditions. Then the wave function satisfies periodic boundary conditions:} 
\begin{eqnarray}
\phi(x+a) &=& \phi(x), 
\label{eq:boundcond}
\end{eqnarray}
\noindent with solutions that are the plane wave states,
\begin{equation}
\phi(x) \sim e^{ikx},
\label{plane_wave}
\end{equation}
\noindent where $k^2 \equiv {2mE}/{\hbar^2}$. Here, $k$ can be either negative or positive.
Imposition of the boundary conditions in Eq.~\ref{eq:boundcond} then requires
\begin{equation}
ka = 2 n \pi,
\label{eq:ka2npi}
\end{equation}
where $n$ is an integer: $n = ...-2, -1, 0, 1, 2, ....$ The eigenvalues are then 
\begin{equation}
E_n = 4 \left( \frac{n^2 \pi^2 \hbar^2}{2ma^2} \right) = 4n^2 E_1^{(0)}.
\label{eq:periodicen}
\end{equation}
Note that these differ from the eigenvalues of an infinite square well: (i) they have a value of $4\times$ 
those eigenvalues, for the same integer, $n$, but
(ii) they are doubly degenerate, and (iii) $E=0$ is possible, and also constitutes the one exception to point (ii) ($n=0$).
The orthonormal basis states are then
\begin{equation}
\phi_n(x) = \sqrt{\frac{1}{a}} \exp{\left[{i\frac{2 \pi n}{a} x}\right]},
\label{eq:repeateigenstate}
\end{equation}
where $n$ is an integer.

\section{HARMONIC OSCILLATOR}

\noindent{The first question one should ask is, does this basis work? And if so, is this basis any better or worse than
the infinite square well basis? To this end we recall the results obtained in Ref.~[\onlinecite{marsiglio09}], for a
harmonic oscillator potential embedded in the infinite square well.}


\subsection{Infinite square well}

\noindent{We placed the harmonic oscillator potential inside an infinite square well of width $a$ that spanned $0<x<a$.
In dimensionless form the potential was written}
\begin{equation}
v_{\text{HO}} = \frac{V_{\text{HO}}}{E_1^{(0)}} = \frac{\pi^2}{4} \left( \frac{\hbar \omega}{E_1^{(0)}} \right)^2 \left( \frac{x}{a} - \frac{1}{2} \right)^2.
\end{equation}
Using this potential, the dimensionless Hamiltonian matrix components are
\begin{eqnarray}\nonumber
h_{nm} &=& \frac{H_{nm}}{E_1^{(0)}} = \delta_{nm} \left[ n^2 + \frac{\pi^2}{48} \left( \frac{\hbar \omega}{E_1^{(0)}}\right)^2 \left( 1 - \frac{6}{(\pi n)^2} \right)\right]  \\ &&
+ (1 - \delta_{nm}) \left( \frac{\hbar \omega}{E_1^{(0)}} \right)^2 g_{mn},
\label{eq:iswhnm}
\end{eqnarray}
where
\begin{equation}
g_{mn} = \left( \frac{(-1)^{n+m} + 1}{4} \right) \left( \frac{1}{(n-m)^2} - \frac{1}{(n+m)^2}\right).
\end{equation}

A diagonalization of this matrix with a truncated basis indeed converged to the correct ground state wave function, and the
coefficients in Eq.~(\ref{expansion}) indeed agreed to high accuracy with those expected from the exact analytical solution:\cite{remark1}
\begin{equation}
c_n = \begin{cases}
i^{(n-1)} \left( \frac{32}{\pi} \frac{E_1^{(0)}}{\hbar \omega} \right)^{1/4} \exp\left[ -n^2 \frac{E_1^{(0)}}{\hbar \omega}\right] & \text{for } n \text{ odd}, \\
0 & \text{for } n \text{ even.}
\end{cases}
\label{eq:hogsiswfourier}
\end{equation}

\subsection{Periodic boundary conditions}

\noindent{With periodic boundary conditions, we use $-{a \over 2} < x < {a \over 2}$, and the potential is given by}
\begin{equation}
V(x) = \frac{1}{2}m\omega^2 x^2 \phantom{aa} {\rm for} \phantom{aa} -\frac{a}{2} < x < \frac{a}{2}
\label{harm_pot_period}
\end{equation}
The required matrix elements are then 
\begin{eqnarray}\nonumber
H_{nm}^V &=& \bra{\phi_n} V(x) \ket{\phi_m} \\
&=& \frac{1}{a} \int_{-a/2}^{a/2} \dif x \, e^{-i \frac{2 \pi nx}{a}} \left( \frac{1}{2}m\omega^2 x^2 \right) e^{i \frac{2 \pi mx}{a} }.
\end{eqnarray}
In dimensionless form this becomes
\begin{equation}
h_{nm}^V \equiv {H_{nm}^V \over E_1^{(0)}} = \left({\hbar \omega \over E_1^{(0)}}\right)^2 \frac{\pi^2}{4} I,
\label{eq:HnmV}
\end{equation}
where
\begin{equation}
I = \int_{-1/2}^{1/2} u^2 e^{i 2\pi (m-n) u} \dif u.
\label{eq:repeatint}
\end{equation}
This integral is straightforward; when combined with the kinetic energy contribution, one obtains
\begin{eqnarray}
h_{nm} &=&  \delta_{nm} \left[ 4n^2 + \frac{\pi^2}{48} \left( \frac{\hbar \omega}{E_1^{(0)}} \right)^2 \right] \nonumber  \\
&+& (1 - \delta_{nm}) \left[ \frac{1}{8} \left( \frac{\hbar \omega}{E_1^{(0)}} \right)^2 \frac{(-1)^{m-n}}{(m-n)^2} \right],
\label{eq:repeathnm}
\end{eqnarray}
which is clearly different from Eq.~\ref{eq:iswhnm}. 

As in the infinite square well case we will have to truncate this matrix, so it makes sense to arrange the basis states
in order of increasing energy. We will use an ordering of quantum numbers as the series $\{ 0, 1, -1, 2, -2, 3, -3, ...\}$, and
then truncate at some $n_{\rm max}$. We can also anticipate our numerical results by calculating 
the Fourier coefficients of the harmonic oscillator ground state analytically. Using the analytical from for the ground state
wave function written in the same dimensionless units as above, 
\begin{equation}
\psi_{\text{HO}}(x) = \left[ \frac{\pi}{2a^2} \left( \frac{\hbar \omega}{E_1^{(0)}}\right) \right]^{1/4} \exp{\left[ -\frac{\pi^2}{4} \left( \frac{\hbar \omega}{E_1^{(0)}}\right) \left({x \over a}\right)^2 \right]},
\label{eq:hogsrep}
\end{equation}
we can take the inner product with each eigenstate from Eq.~\ref{eq:repeateigenstate}. We obtain
\begin{eqnarray}
c_n &=& \braket{\phi_n(x) \psi_{\text{HO}}(x)} \nonumber \\
&= &\left[ \frac{8}{\pi} \left(  \frac{E_1^{(0)}}{\hbar \omega}  \right)\right]^{1/4} \exp{\left[ -4n^2 \left( \frac{E_1^{(0)}}{\hbar \omega}  \right)\right]}
\label{eq:repeatcoeff}
\end{eqnarray}
which again differ from those in Eq.~(\ref{eq:hogsiswfourier}).

\section{NUMERICAL COMPARISON}

\begin{figure}
\includegraphics[width=0.5\textwidth]{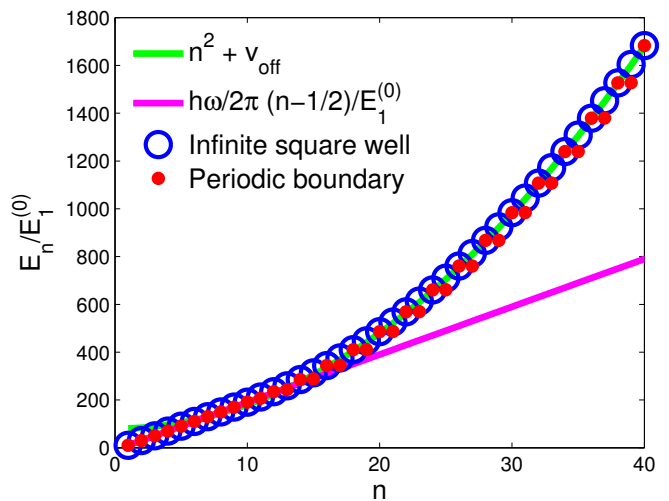}
\caption{Plot of the normalized energy levels vs. quantum number $n$ for numerical solutions using open (infinite square well)
and periodic boundary conditions. We used $n_{\rm max} = 60$ and $\hbar \omega / {E_1^{(0)} }= 20$, and note that
our quantum number $n$ begins at unity (not zero). The rationale for the curve $n^2 + v_{\rm off}$ is explained in Ref. [\onlinecite{marsiglio09}].
Both results give the correct energy eigenvalues at low values of $n$, and both grow as $n^2$ at large values of $n$, reflecting the expected behaviour due to the confining box. Note the clear degeneracies at large $n$ for the case with periodic boundary condition energies.}
\label{fig:eiswvserep}
\end{figure}

\noindent{We pause for a moment to compare the accuracy and efficiency of the two basis sets. In Fig.~\ref{fig:eiswvserep} we plot the numerical eigenenergies obtained for the two cases with $n_{\rm max} = 60$. Both methods give very similar results.}

\begin{figure}
\includegraphics[width=0.5\textwidth]{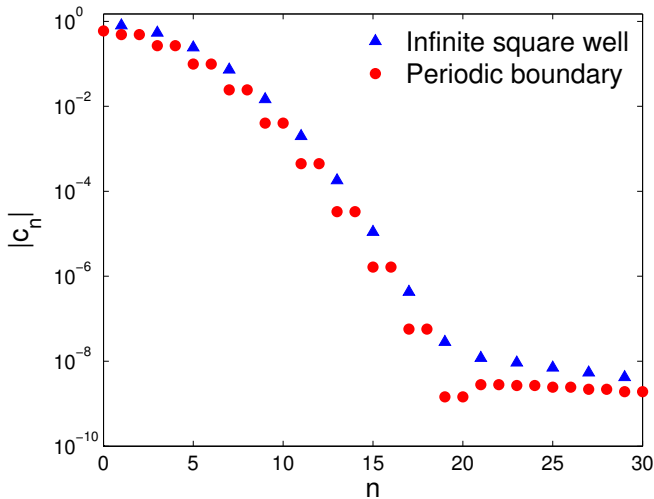}
\caption{The numerically-derived absolute value of the Fourier coefficients for the infinite square well and the periodic boundary condition embedding potentials, plotted on a semilog graph. The even cases for the infinite square well have finite nonzero results on the order of less than $10^{-15}$ and are not shown. Here, the truncated matrix dimension is $n_{\rm max}=60$ and $(\hbar \omega / E_1^{(0)}) = 20$. Analytical results cannot be distinguished from the numerical results.}
\label{fig:cnlog}
\end{figure}

In particular, either result reproduces the expected linear in $n$ results to high accuracy for low lying states, 
where the presence of the box
is not even noticed. This is the regime that is physically relevant to the harmonic oscillator potential.
At higher values of $n$ both cases cases produce energies that grow as $n^2$. To evaluate the
efficiency, we show the results for the coefficients 
on a log plot in Figure \ref{fig:cnlog}; both methods yield similar results, with the smooth decay 
breaking down where $n \approx \hbar \omega / E_1^{(0)}$, which is the condition for the basis state energy  equal
to the crossover (from harmonic oscillator to square well) energy for the potential. Exact analytical results agree to 7
digits in either case. Note that agreement can be systematically improved indefinitely in both cases by enlarging 
the width of the embedding square well.

\section{BLOCH'S THEOREM}

\noindent{Bloch's Theorem\cite{bloch29} states that the electronic wave function in a periodic potential can be written as a
product of a plane wave and a function which has the same periodicity as the potential. Alternatively one can write
\begin{equation}
\psi(x+a) = e^{iKa} \psi(x)
\label{eq:blochstheorem}
\end{equation}
where $a$ is the unit cell length and $K$ is a wave vector such that $-\pi < Ka < \pi$. 

\subsection{Analytical solution to the Kronig-Penney model}

\noindent{In practice, the analytical treatment
of the Kronig-Penney model\cite{kronig31} allows one to utilize Bloch's Theorem as a boundary condition. In that problem,
one uses a potential that repeats a square well of width $b$ followed by a square barrier of width $a-b$:}
\comment{
\begin{equation}
V(x) = \begin{cases}
0 & 0 < x < b, \\
V_0 & b < x < a.
\end{cases}
\label{periodic_potential}
\end{equation}
}
\begin{equation}
V(x) = V_0 \sum_{n=-\infty}^{+\infty} \theta \left[ x - \left(na + b \right) \right]  \theta \left[ \left(n+1 \right) a -x \right] ,
\label{periodic_potential}
\end{equation}
\noindent where $\theta[x]$ is the Heaviside step function.
Here, the overall repeat distance is $a$. For states whose energies are below the barrier height one can solve the
Schr\"odinger equation for each region in turn: a linear combination of two plane waves in the `well' regions, followed by
a linear combination of an exponentially decaying and an exponentially growing solution in the `barrier regions. Matching
the wave functions and their derivatives at the interface determines two of the unknown coefficients; at the next interface,
a similar procedure determines two more coefficients, but this still leaves two unknown coefficients, and on this would
go ad infinitum. Bloch's Theorem allows the second matching process to terminate the procedure, since now the two
coefficients are written in terms of the original two, and we are left with four homogeneous equations with four unknowns. 
Since the determinant of the coefficients of these four equations must be zero for there to be a solution, this results in the 
expression
\begin{eqnarray}
\cos{(Ka)} &=& \cos{(k_1 b)} \cosh{[\kappa_2 (a-b)]} \nonumber \\ 
&+& \frac{\kappa_2^2 - k_1^2}{2k_1 \kappa_2} \sin{(k_1 b)} \sinh{[\kappa_2 (a-b)]}
\label{eq:kpanalsol}
\end{eqnarray}
where $k_1 = \sqrt{2mE/\hbar^2}$ and $\kappa_2 = \sqrt{2m(V_0-E)/\hbar^2}$. Equation~(\ref{eq:kpanalsol}) is thus
an implicit equation for $E(K)$. In practice, one selects a value of $E$; the absolute value of the right-hand-side of
Eq.~(\ref{eq:kpanalsol}) is evaluated --- it is either greater or less than unity. If greater, then there is clearly no 
solution possible, while
if less than unity, then taking the inverse cosine of this quantity gives the value of $Ka$ for which this energy is the
solution. Plots will be shown later when comparisons are made with the numerical results.

\subsection{Matrix method for the Kronig-Penney model}

\noindent{It should be clear from Eq.~(\ref{eq:blochstheorem}) why we altered the original matrix method to include the case
of periodic boundary conditions; now we simply replace Eq.~(\ref{eq:boundcond}) with Eq.~(\ref{eq:blochstheorem}).
How does this alter the procedure discussed in the previous section? Quite simply, Eq.~(\ref{eq:ka2npi}) is replaced
by the one required by Bloch's Theorem:}
\begin{equation}
ka = 2\pi n    \phantom{aaaa} \rightarrow  \phantom{aaaa} ka - Ka = 2 \pi n
\label{bloch_periodic}
\end{equation}
where, as before, $k^2 = {2mE}/{\hbar^2}$. Hence
\begin{eqnarray}\nonumber
E_n^{(0)} &=& \frac{\hbar^2 \pi^2}{2ma^2} \left( 2n + \frac{Ka}{\pi} \right)^2 \\
&=& \, \, \, \, E_1^{(0)} \left( 2n + \frac{Ka}{\pi} \right)^2.
\label{eq:blochmodify}
\end{eqnarray}
Crucially, Eq.~\ref{eq:blochmodify} applies only to the diagonal elements; the 
off-diagonal elements are unaffected, because the basis states are modified just as
\begin{eqnarray} \nonumber
\phi_n(x) &=& \frac{1}{\sqrt{a}} e^{i\frac{2\pi n}{a} x} \rightarrow \frac{1}{\sqrt{a}} e^{i\frac{2\pi n + Ka}{a} x} \\
&=& \frac{1}{\sqrt{a}} e^{i\frac{2\pi n}{a} x} e^{+iKx} = e^{+iKx} \phi_n(x).
\end{eqnarray}
But this means that $\phi_n^* \rightarrow e^{-iKx}\phi_n^*$. Thus the extra exponentials arising from the Bloch condition
cancel one another when we calculate $H_{nm}^V$ and so the off-diagonal terms remain purely potential dependent. 

To generate bands of energy, different values of $Ka$ in the region $(-\pi, \pi)$ are passed to matrices suitably modified by Eq.~\ref{eq:blochmodify}.
The remaining matrix elements are determined as before. For the case of a well of width $b$ centred between two barriers
of height $V_0$, each of width $(a-b)/2$ [for a periodic array, this is just a shifted version of Eq.~(\ref{periodic_potential})---see Fig.~(\ref{singlewell_central})], then,
with $\rho \equiv b/a$ and $v_0 \equiv V_0/E_1^{(0)}$, we have
\comment{
\begin{eqnarray}
h_{nm} &=& \left[ 4n^2 - \frac{4n (Ka)}{\pi} + \frac{(Ka)^2}{\pi^2} + v_0 (1-\rho) \right] \delta_{nm} \nonumber \\
& & -\frac{iv_0}{2\pi (m-n)}  \left[ e^{-i 2\pi (m-n) (\frac{1}{2}-\frac{\rho}{2})} \right. \\ \nonumber
& & -\left. e^{-i 2\pi (m-n) (\frac{1}{2}+\frac{\rho}{2})} \right] (1-\delta_{nm}).
\label{eq:periodicstepblochcentered}
\end{eqnarray}
}
\begin{eqnarray}
h_{nm} &=& \delta_{nm} \left[ \left( 2n + \frac{Ka}{\pi} \right)^2 + v_0 (1-\rho) \right] \\
& &+(1-\delta_{nm}) v_0 {(-1)^{m-n+1} \over \pi} {\sin{[\pi(m-n) \rho]} \over m-n}. \nonumber
\label{eq:periodicstepblochcentered}
\end{eqnarray}

We repeatedly diagonalize matrices of this form for various values of $Ka \in (-\pi, \pi)$ to form the band solutions.

\begin{figure}[h]
\includegraphics[width=0.5\textwidth]{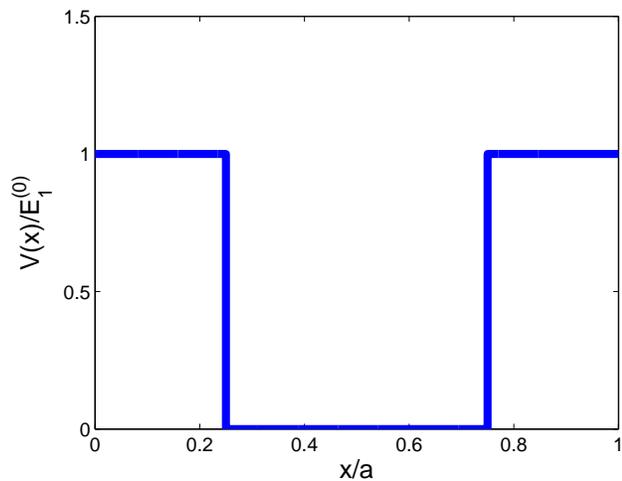}
\caption{Central square well with $v_0 = 1$ and $\rho = 0.5$.}
\label{singlewell_central}
\end{figure}

\begin{figure}
\subfloat{\includegraphics[width=0.5\textwidth]{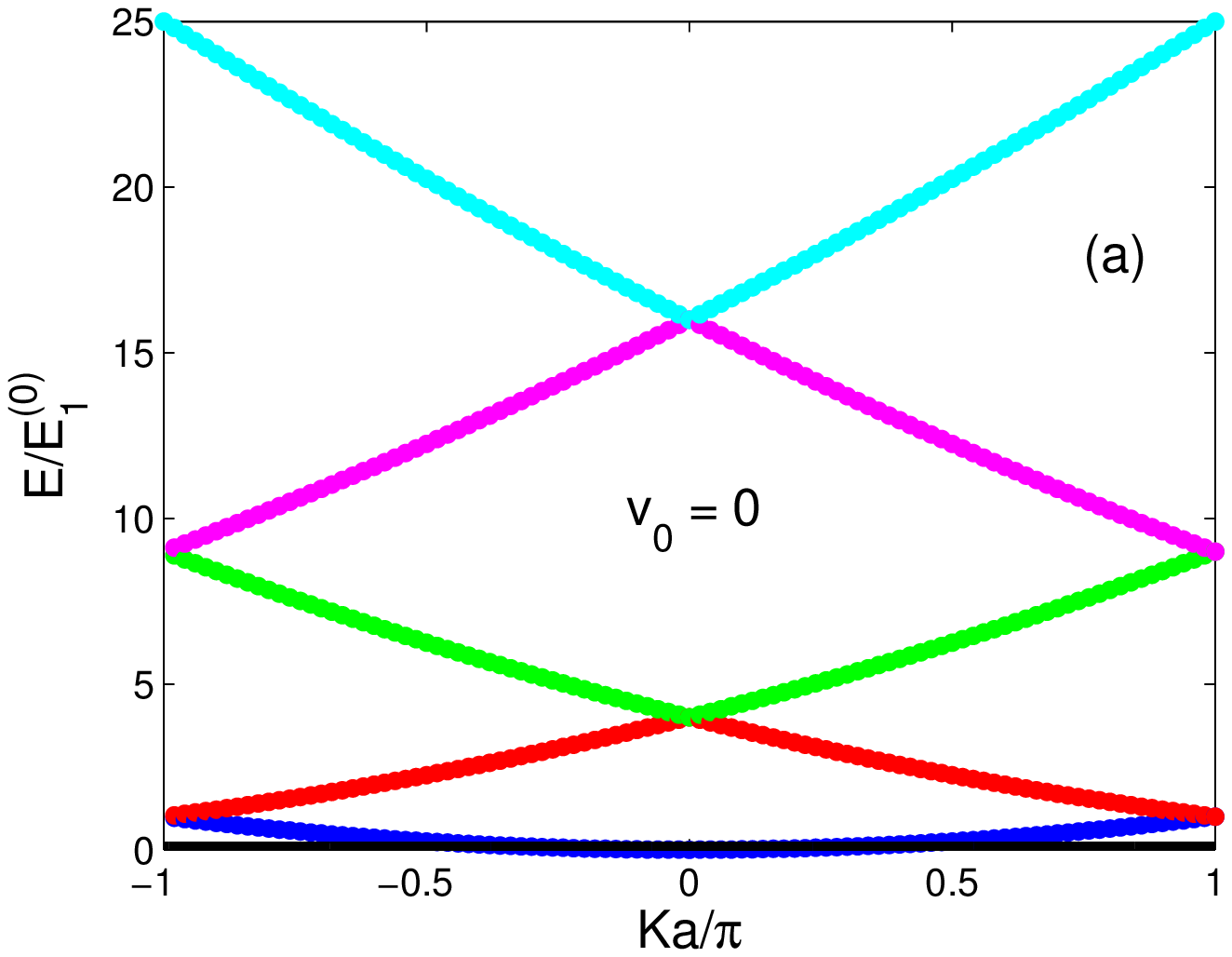}} \label{fig:kpv0} \\
\subfloat{\includegraphics[width=0.5\textwidth]{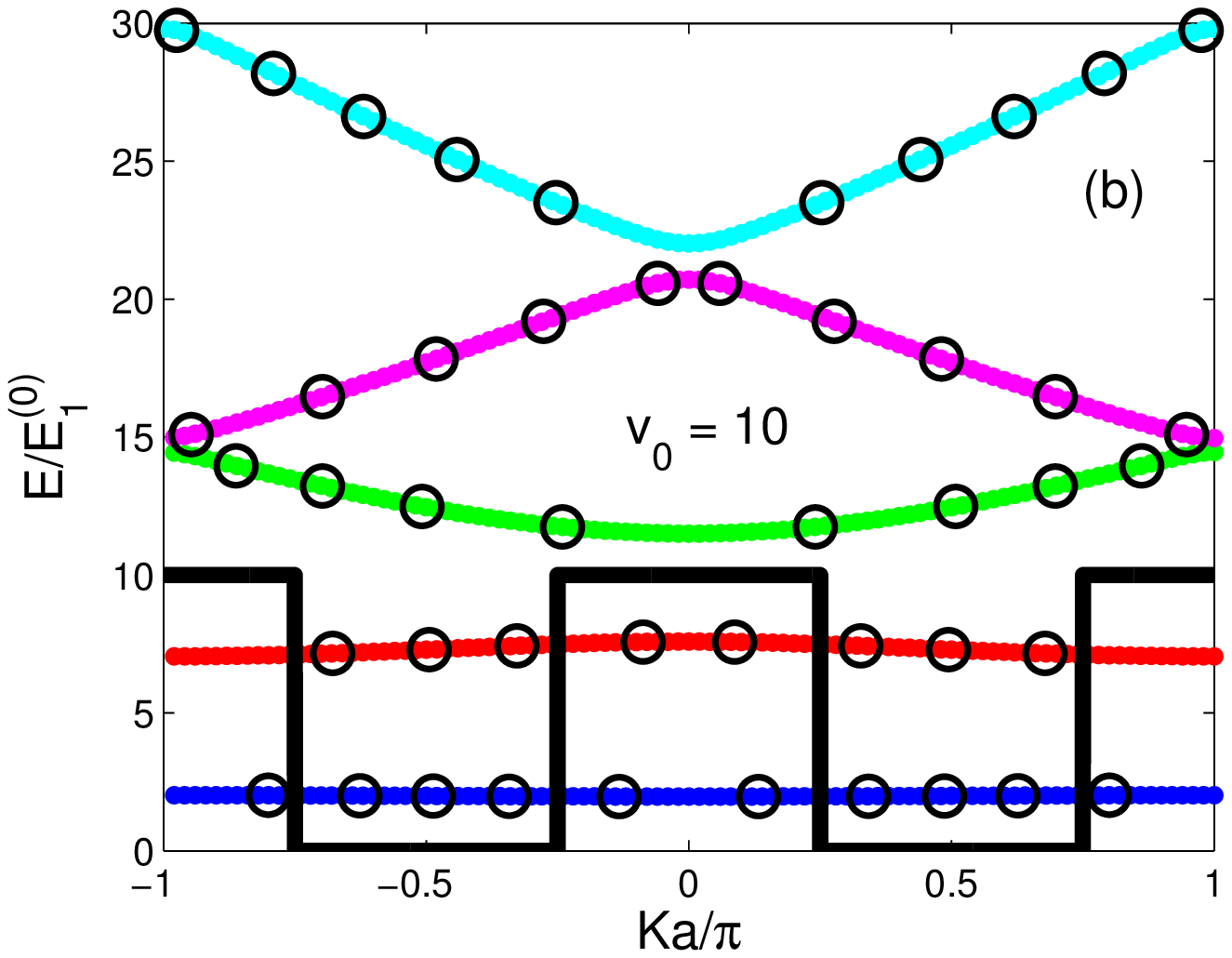}} \label{fig:kpanal}
\caption{Solutions to the Kronig-Penney potential with (a) no well/barrier, and (b) a repeated well/barrier sequence with
barrier height $v_0 = 10$. In (b) we have also indicated with open circles some of the analytical solutions to Eq.~(\ref{eq:kpanalsol}). A schematic representation of the potential is shown to indicate the boundedness of the energy bands.
The matrix solutions utilize $n_{\rm max} = 60$, and we have used $\rho = b/a = 0.5$.}
\label{fig:kronigpenneygraphs}
\end{figure}

Some results for this model are shown in Fig.~\ref{fig:kronigpenneygraphs}. When there are no barriers ($v_0 = V_0/E_1^{(0)} = 0$) we see the folded parabolas, corresponding to plane wave energy solutions. As we turn the periodic potential on, e.g.~with
$v_0 = 10$, we see bandgaps emerge, with larger gaps in the lower energy bands, and more curvature (i.e. bandwidth) in the upper bands. Analytical solutions found by solving Eq.~\ref{eq:kpanalsol} are overlaid with the numerical ones, and are in
complete agreement. We have thus succeeded in demonstrating that the numerical method works for periodic arrays.

\subsection{General potential shapes}

\noindent{It should be clear that we now have freedom to study periodic arrays of potentials with any shape, simply by
performing many calculations (corresponding to different values of $Ka$) for the wave function {\em in one unit cell}.}
\begin{figure}
\includegraphics[width=0.5\textwidth]{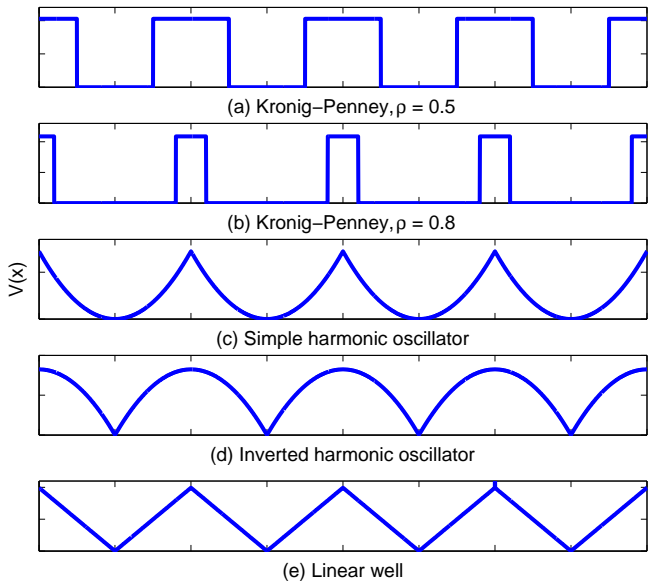}
\caption{Schematic representation of the potentials used for comparing energy band structures.}
\label{fig:potentials}
\end{figure}
See Fig.~\ref{fig:potentials} for various examples of periodic potentials, but note that they do not necessarily 
have to have an analytical
form, nor do they necessarily require an analytical integration for the matrix elements. 
For example, we can repeat harmonic oscillator wells. Then the required dimensionless matrix elements are
\begin{eqnarray}
h_{nm} &=& \delta_{nm} \left[ \left( 2n + \frac{Ka}{\pi} \right)^2 + \frac{\pi^2}{48} \left( 
\frac{\hbar \omega}{E_1^{(0)}} \right)^2 \right]  \nonumber \\
&+& (1 - \delta_{nm}) \left[ \frac{1}{8} \left( \frac{\hbar \omega}{E_1^{(0)}} \right)^2 \frac{(-1)^{m-n}}{(m-n)^2} \right].
\label{eq:periodicshobloch}
\end{eqnarray}

For the inverted harmonic oscillator potential (see Fig.~\ref{fig:potentials})
we use
\begin{equation}
V(x) = \begin{cases}
-\frac{1}{2}m\omega^2 \left[ x^2 - \frac{a^2}{4} \right] & 0 < x < \frac{a}{2}, \\
-\frac{1}{2}m\omega^2 \left[ (x-a)^2 - \frac{a^2}{4} \right] & \frac{a}{2} < x< a.
\end{cases}
\end{equation}
The matrix elements are then
\begin{eqnarray}\nonumber
h_{nm} &=& \delta_{nm} \left[ \left( 2n + \frac{Ka}{\pi} \right)^2 + \frac{\pi^2}{24} \left( \frac{\hbar \omega}{E_1^{(0)}} \right)^2 \right] \nonumber \\
&-&  (1 - \delta_{nm}) \left[ \frac{1}{8} \left( \frac{\hbar \omega}{E_1^{(0)}} \right)^2 \frac{(-1)^{m-n}}{(m-n)^2} \right].
\label{eq:invertedperiodicshobloch}
\end{eqnarray}
Note the difference between the above equation and Eq.~\ref{eq:periodicshobloch} in both the off-diagonal elements (change of sign) and the diagonal elements (factor of 2 in the potentlal term).
%

A third example is the linear well (resembles a saw-tooth in Fig.~\ref{fig:potentials}), with potential defined by
\begin{equation}
V(x) = \begin{cases}
2A\left(\frac{1}{2} - \frac{x}{a} \right) & 0 < x < \frac{a}{2}, \\
2A\left(\frac{x}{a} - \frac{1}{2} \right) & \frac{a}{2} < x < a.
\end{cases}
\end{equation}
This has matrix elements
\begin{eqnarray}
h_{nm} &=& \delta_{nm} \left[ \left( 2n + \frac{Ka}{\pi} \right)^2 + \frac{A}{2}  \right] \nonumber \\
&-& (1 - \delta_{nm}) \frac{A}{\pi^2(m-n)^2} \left[1 - (-1)^{m-n}\right].
\label{eq:linearbloch}
\end{eqnarray}

As mentioned earlier, potentials can also be utilized for which the matrix elements have no known analytical solution, or in which obtaining the analytical solution would be overly cumbersome.
One such potential is the so-called ``pseudo-Coulomb" potential which has the form
\begin{equation}
v(x) = \frac{V(x)}{E_1^{(0)}} = \frac{-A}{\sqrt{(x-\frac{a}{2})^2 + b^2}}
\label{eq:pseudocoulomb}
\end{equation}
with $A$ a positive number representing the strength, or alternatively, the
inverted pseudo-Coulomb potential with $A$ negative, and $b$ is a small numerical factor introduced to prevent singularities (the true Coulomb potential is recovered as $b \rightarrow 0$).

\subsection{Comparing band structures}

\noindent{Results for the band structures corresponding to the potential shapes in Fig.~\ref{fig:potentials} are shown in Figs.~\ref{fig:bandskpv05}, \ref{fig:bandskpv08}, \ref{fig:bandssho}, \ref{fig:bandsinvsho}, \ref{fig:bandslin}. We chose
parameters such that in all cases three bands with energies less than the maximum barrier height would form, and in which 
the third band would have an energy difference between the highest level (at $Ka  = \pm \pi$) and the maximum barrier potential, $V_\text{max}$, of $\Delta E =  E_1^{(0)}$ in each case.} 

Note that bands with gaps form at energies above the barrier maxima as well. However, the character of these bands is
strongly dependent on the type of periodic potential used. In particular, for the latter three potentials the gap can be quite
small (not discernible, for example, on the scale of Fig.~\ref{fig:bandsinvsho}). Moreover for these latter three potentials,
the minima and maxima of these higher energy bands can be distinctly non-parabolic, and in fact exhibit V-shaped (or
inverted V-shaped) dispersions close to the minima (or maxima).

\begin{figure}
\includegraphics[width=0.5\textwidth]{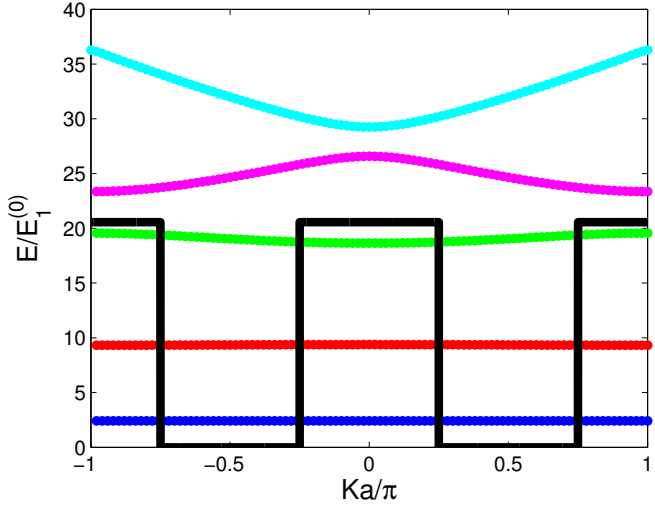}
\caption{Energy band diagram for the Kronig-Penney potential with $\rho = 0.5$ and $v_0 = 20.5607$.
Note the presence of energy gaps for the high energy bands. These gaps persist with diminishing size
as the energy increases.}
\label{fig:bandskpv05}
\end{figure}

\begin{figure}
\includegraphics[width=0.5\textwidth]{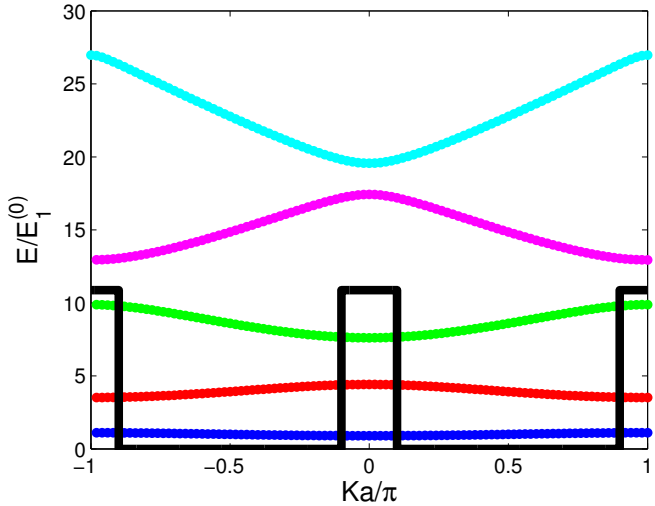}
\caption{Energy band diagram for the Kronig-Penney potential with $\rho = 0.8$ and $v_0 = 10.8775$.}
\label{fig:bandskpv08}
\end{figure}

\begin{figure}
\includegraphics[width=0.5\textwidth]{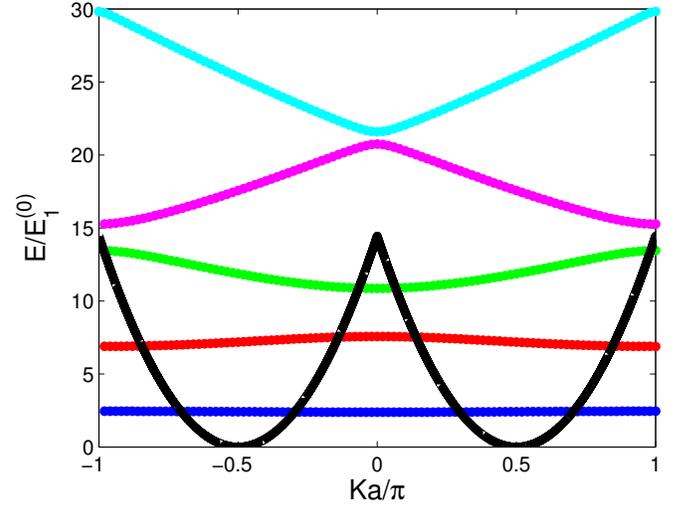}
\caption{Energy band diagram for the simple harmonic oscillator potential with $\hbar\omega/E_1^{(0)} = 4.84105$.
Note the small energy gap separating the two highest energy bands (compared with that arising from
the Kronig-Penney potential shown in the previous two figures). This gap is even smaller for the inverted
harmonic oscillator potential (see Fig.~\ref{fig:bandsinvsho}) and the linear well potential (see Fig.~\ref{fig:bandslin}).}
\label{fig:bandssho}
\end{figure}

\begin{figure}
\includegraphics[width=0.5\textwidth]{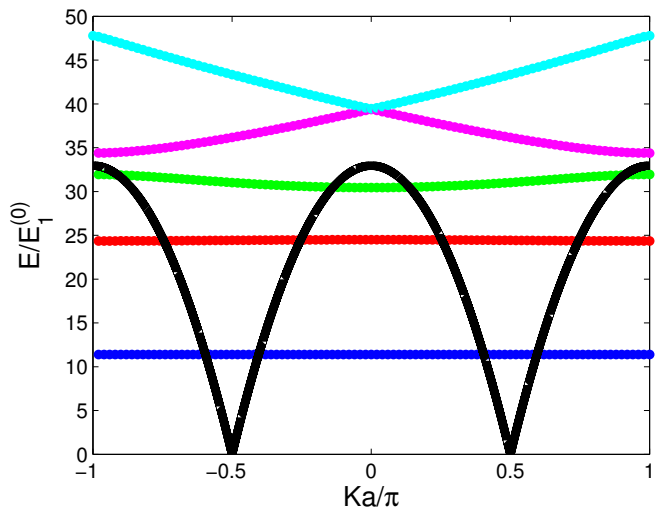}
\caption{Energy band diagram for the inverted harmonic oscillator potential with $\hbar\omega/E_1^{(0)} = 7.30845$.
Note the very small gap between the two highest energy bands, and also the non-parabolic dispersion for these
two bands.}
\label{fig:bandsinvsho}
\end{figure}

\begin{figure}
\includegraphics[width=0.5\textwidth]{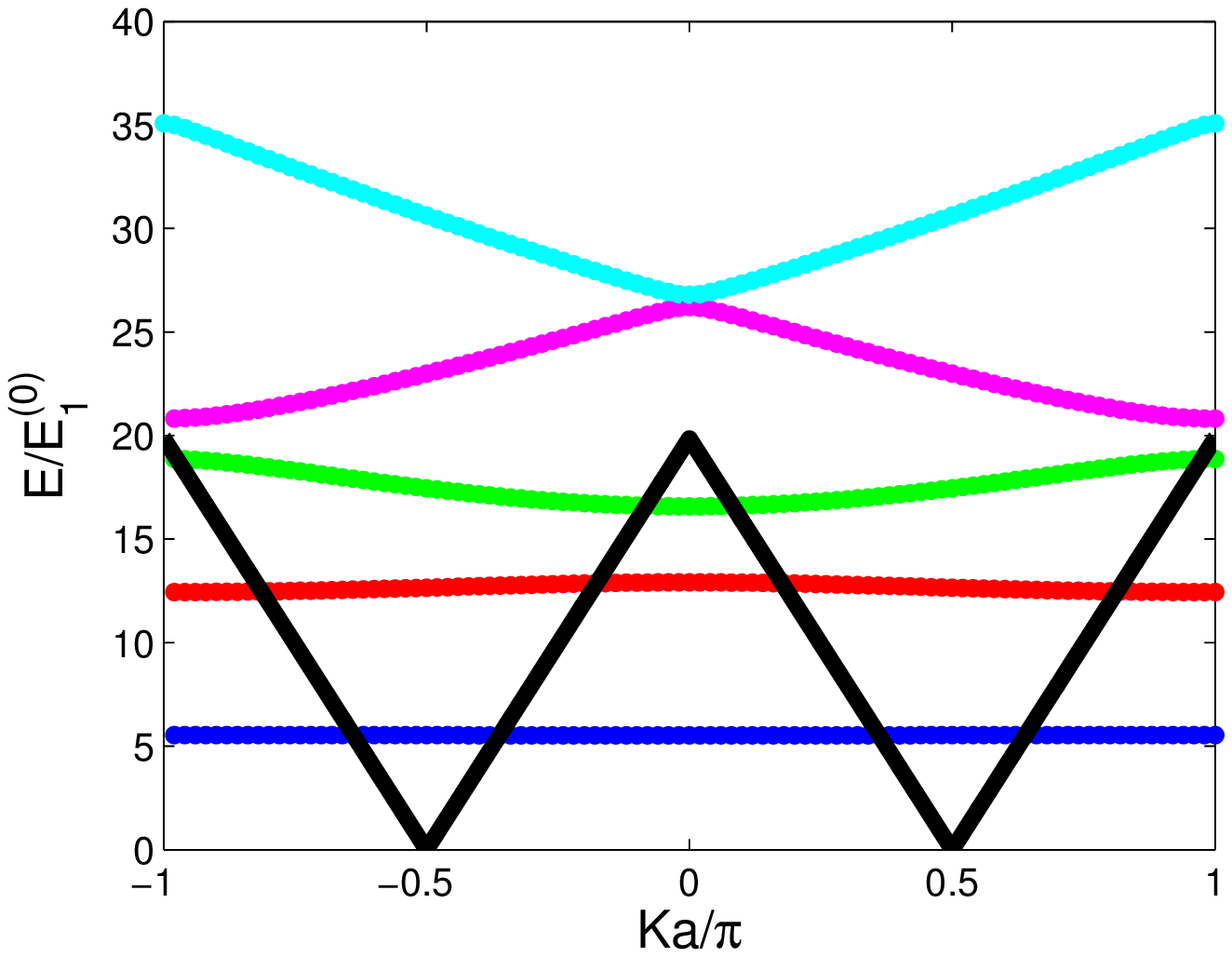}
\caption{Energy band diagram for the linear well potential with $A = 19.8705$. Comments from the previous two figures
apply here as well.}
\label{fig:bandslin}
\end{figure}

We then focused on the third band (the one closest to but lower than $V_\text{max}$ in each case and normalized the bands by setting the highest band level to $E = 0$ in each case, as shown in Fig.~\ref{fig:thirdbandsubtracted}. 
\begin{figure}
\includegraphics[width=0.5\textwidth]{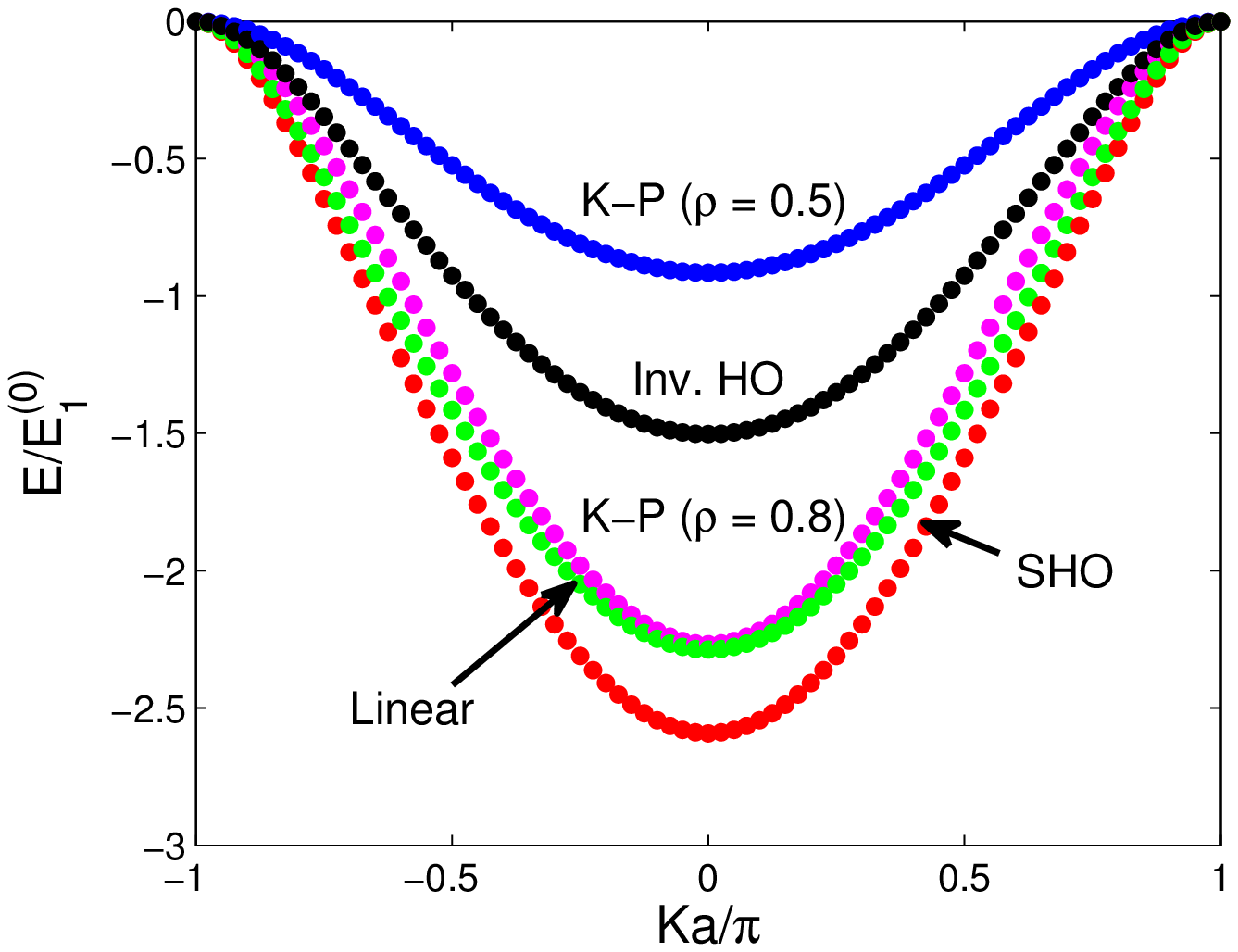}
\caption{The third energy band for the five cases in Fig.~(\ref{fig:potentials}), with the topmost energy set equal to zero.
Note the variation in bandwidth, but, more importantly, the difference in curvatures, as tabulated in Table III.}
\label{fig:thirdbandsubtracted}
\end{figure}
The bandwidth varies considerably as shown; this can be adjusted by varying the barrier heights and widths, as is explicitly
shown in the case of the Kronig-Penney model (first two figures in Fig.~\ref{fig:potentials}); in the case of the other
potentials we could include an additional barrier or well `plateau'. One property of importance, however, is the ratio
of the electron and hole effective masses. These are defined through the correspondence of the band dispersion with
the free-electron-like parabolic behaviour at both the minima and maxima of the bands. Hence
\begin{equation}
{1 \over m^\ast_{\rm ele \atop hol}} = {1 \over \hbar^2}\frac{\dif^{\, 2} \epsilon}{\dif^{\, 2} K} \biggl{|}_{K_{\rm min} \atop K_{\rm max}}.
\label{mass}
\end{equation}
The second derivative in Eq.~(\ref{mass}) is evaluated with a five-point fit,\cite{pang2006} and the ratio of the
two effective masses for each potential shape is tabulated in Table III. The two Kronig-Penney models yield
essentially the same ratio, and the magnitude of this ratio can decrease considerably when other potential
shapes are considered. First note that the ratio has a magnitude that is less than unity. This is because holes have
effectively a weaker barrier through which to tunnel, compared with electrons. This decrease is most pronounced when
either the linear or simple harmonic oscillator potential is used. The reason is that these have cusp-like barriers, so that the barrier width is also reduced for holes compared to electrons; therefore the holes should have more mobility (i.e. lower effective
mass) compared with electrons, over and above the advantage already present due to the difference in effective
barrier height. This asymmetry concerning electrons and holes may be important in a new class of superconducting models,
where dynamic interactions are taken into account.\cite{hirsch01,bach10}

\begin{table}
\centering
\caption{The second derivates at the peak and trough of the third energy bands from Fig.~\ref{fig:thirdbandsubtracted}, along with the ratio of $\epsilon_\text{ele}''/\epsilon_\text{hol}''  \equiv m_{\rm hol}^\ast/m_{\rm ele}^\ast $.}
\begin{tabular*}{0.49\textwidth}{@{\extracolsep{\fill}} l r r r}
\toprule[1.5pt]
Potential			&	$\epsilon_\text{ele}''$	&	$\epsilon_\text{hol}''$	&	$\epsilon_\text{ele}''/\epsilon_\text{hol}''$ \\
\toprule[1.5pt]
K-P ($\rho = 0.5$)	&	13.83   		&	-25.35		&	-0.55 \\
K-P ($\rho = 0.8$)	&	39.09			&	-70.61		&	-0.55 \\
Simple HO		    &	37.84   		&	-121.80		&	-0.31 \\
Inverted HO		    &	19.83			&	-55.96		&	-0.35 \\
Linear			    &	31.63   		&	-102.23		&	-0.31 \\
\bottomrule[1.5pt]
\end{tabular*}
\label{tab:secondderivatives}
\end{table}

\section{Summary}

\noindent{We hope that this paper will serve as a bridge from the classroom to the research desk, for both undergraduate
and graduate students, in the area of electronic structure calculations. The `classroom' or `textbook' example for
electronic band structure has always been the Kronig-Penney model. The enlightening feature in the model has
been the recasting of the problem of an extended (i.e. Bloch) state in terms of the problem in one unit cell, due
to Bloch's Theorem. The part that is more intimidating for students is the extension beyond simple square wells
to more realistic potentials; this is what this paper has tried to address.}

Simple procedures to obtain numerical solutions for any potential that supports bound states were illustrated in
Ref.~[\onlinecite{marsiglio09}]. This methodology was extended to periodic boundary conditions for the purpose
of this paper, because the next step, implementing the Bloch condition, is then essentially trivial. Students can
then tackle their own favourite potential, for example the ``pseudo-Coulomb" potential referred to in 
Eq.~(\ref{eq:pseudocoulomb}). We have given a number of examples already in this paper, and mentioned one
particular property---the asymmetry between electron and hole masses, which can have important consequences 
for semiconductors and possibly superconductors. In general the more realistic potentials tend to make the barrier
effectively narrower for hole than for electrons, with the consequence that hole masses are generally lower
than electron masses. We also note that more realistic potentials can give rise to higher energy bands with non-parabolic
shapes.

\begin{acknowledgments}

\noindent{This work was supported in part by the Natural Sciences and Engineering Research Council of Canada (NSERC), by the Alberta iCiNano program, and by a University of Alberta Teaching and Learning Enhancement Fund (TLEF) grant.}

\end{acknowledgments}


\begin{thebibliography}{99}
\bibitem{remark1} See Ref. [\onlinecite{marsiglio09}]. Note that this reference had a typo in Eq. (15), which has
been corrected here in Eq.~(\ref{eq:hogsiswfourier}).
\bibitem{remark2} In either case, for the harmonic oscillator potential, or any potential that is even in $x$, we could
of course increase the efficiency by a factor of two, by leaving out the even integer basis functions for the infinite square
well case (since they are zero) and using only the sum of the positive and negative coefficients for the periodic boundary
condition case. In a more general case neither of these efficiencies is possible.
\end{thebibliography}
\end{document}